**What do we mean when we say we are clustering multimorbidity?**

Sohan Seth[1], Nazir Lone[2], Niels Peek[3], Bruce Guthrie[2]

[1]School of Informatics, University of Edinburgh,

[2]Usher Institute, University of Edinburgh,

[3]The Healthcare Improvement Studies Institute (THIS Institute), Department of Public Health and Primary Care, University of Cambridge

Corresponding author: Dr Sohan Seth, Lead Data Scientist, School of Informatics, University of Edinburgh

Sohan.Seth@ed.ac.uk

Abstract

Clustering multimorbidity has been a global research priority in recent years. Existing studies usually identify these clusters using one of several popular clustering methods and then explore various characteristics of these clusters, e.g., their genetic underpinning or their sociodemographic drivers, as downstream analysis. These studies make several choices during clustering that are often not explicitly acknowledged in the literature, e.g., whether they are clustering conditions or clustering individuals, and thus, they lead to different clustering solutions. We observe that, in general, clustering multimorbidity might mean different things in different studies, and argue that making these choices more explicit and, more importantly, letting the downstream analysis, or the purpose of identifying multimorbidity clusters, guide these choices, might lead to more transparent and operationalizable multimorbidity clusters. In this study, we discuss various purposes of identifying multimorbidity clusters and build a case for how different purposes can justify the different choices in data and methods.

Introduction

Multimorbidity is the occurrence of two or more long-term conditions in an individual simultaneously[1]. Multimorbidity is increasingly common globally because of rapid population ageing, creating challenges for all healthcare systems, which remain primarily organised around single diseases[2]. It matters because it is associated with increased cost of care, higher mortality and poor quality of life. Clustering long-term conditions or individuals with long-term conditions in 'meaningful' groups has the potential to enrich our understanding of disease trajectories (e.g., how individuals accrue long-term conditions over a lifetime and the genetic[3] and environmental causes of different patterns of accrual) and their impact (i.e., how different trajectories affect outcomes like mortality, quality of life[4], functional status and healthcare resource utilization[5]) at both individual and population level. Such understanding could improve how we target the prevention of multimorbidity or interventions to mitigate the consequences of multimorbidity.

Although exploring multimorbidity clusters has received significant attention in the literature and has been identified as a research priority[6], we observe that it is not always clear what is meant by 'multimorbidity clusters' or 'clustering long-term conditions' because long-term conditions can co-occur through multiple mechanisms including: (1) By chance (which may be more important in older people in whom individual conditions usually have the highest prevalence); (2) Because conditions share a common aetiology (like smoking or obesity or genetic underpinnings); (3) Because one condition causes another (such as diabetes and micro- or macro-vascular disease); (4) Because the treatment for one condition causes another (such as aspirin prescribed for coronary heart disease causing stroke due to cerebral haemorrhage); or (5) Conditions can also be said to cluster if they each lead to a common adverse outcome or care needs, for example, stroke and Parkinson's Disease may be aetiologically unrelated but may still both contribute multiplicatively to functional disability and reduced quality of life. **Figure 1** illustrates these mechanisms.

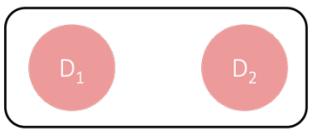
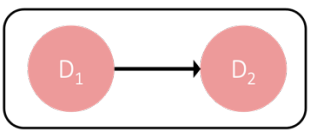
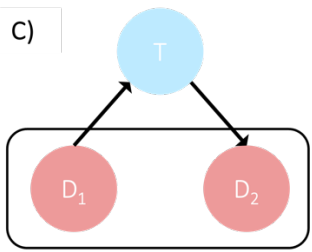
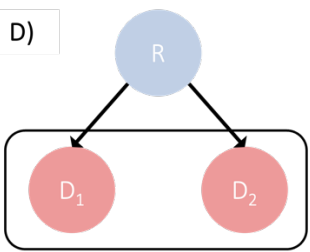
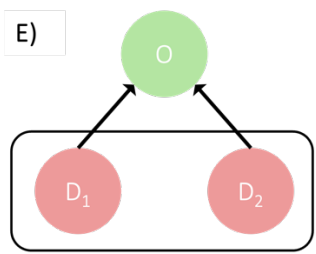

$D_1$ and $D_2$ are a cluster of interest because,
A) They appear frequently by chance
B) $D_1$ leads to $D_2$
C) $D_2$ is an effect of treating $D_1$ by T
D) $D_1$ and $D_2$ have a common risk R
E) $D_1$ and $D_2$ together cause O

*Figure 1: Different situations where finding clusters of multimorbidity can be of interest.*

The concept of clustering is well defined in the context of machine learning and specifically unsupervised learning. It involves dividing objects into coherent groups where objects within a group are more alike than objects in different groups. However, similarity between objects can be defined in different ways, leading to different clustering solutions. For example, *hierarchical clustering* or *k-medoids clustering* require defining a distance metric among objects, and one can choose the metric among various options such as correlation, Jaccard distance, etc. In the context of exploring multimorbidity clusters, analysts always have to make several choices including whether: (1) Clustering is applied over conditions (such as 'asthma, eczema, hay fever') or over individuals (such as 'people with atopy'); (2) Clustering is done on a cross-section of data (presence or absence of conditions at a moment in time) or in longitudinal data (pattern of condition accrual over time); (3) Clustering focuses on joint prevalence of conditions (how often they co-occur) or their strength of association (co-occurrences beyond chance); (4) Clusters are made exclusive (type 2 diabetes is only allowed to be in one condition cluster; or an individual is uniquely allocated to a cluster) or objects can be assigned simultaneously to multiple clusters (type 2 diabetes is allowed to be in clusters of conditions caused by obesity, but also in clusters of conditions that it causes; an individual can simultaneously be allocated to an atopy cluster and a cardiometabolic cluster).

Our primary observation is that in the current literature, the term 'multimorbidity cluster' is often used to mean different things, and the underlying assumptions and applications are not always explicit in published papers. We argue that how researchers define multimorbidity clusters should depend on how these clusters are intended to be used in practice. The various downstream objectives may favour and justify different ways of forming multimorbidity clusters. We suggest that there is no *one* single approach to how multimorbidity clusters *should* be formed. The various approaches available have different rationales and implications, and elaborating on these choices will help the research community appreciate the different approaches to clustering long-term conditions. In this paper, we start by describing the potential applications of clustering in multimorbidity research, followed by common choices made during clustering, and we make the case for the importance of making these choices informed and explicit.

### What is clustering analysis in the context of multimorbidity?

Given a set of objects, a clustering method divides the data into groups or clusters such that objects within a cluster are more similar to each other than objects belonging to two different clusters. **Figure 2**A shows an example where four objects are being clustered in two groups. Each object has a shape and a colour, and in this situation, we can either divide the objects based on their colour or their shape. Depending on context and purpose, both of these groupings could be equally valid and equally relevant for the intended purpose. For example, in the context of multimorbidity, individuals can be grouped based on the number of long-term conditions they have accrued, or the presence and absence of specific conditions, or the body system the long-term conditions are related to (respiratory, cardiovascular, etc.), or the effect of long-term conditions on quality of life. **Figure 2**B illustrates some of these scenarios, e.g., whether we are interested in the order in which the conditions are accrued or not. Additionally, existing studies make several (usually implicit) assumptions for the clustering approach, the choice of domain (cross-sectional or longitudinal), the objects to be clustered (clustering conditions or individuals), the manner of assignment (assigning exclusively or non-exclusively), and the mechanism of interest (prevalence or association). We observe that these choices are often not stated explicitly in published studies, while they can lead to different meanings of 'multimorbidity clusters'. For example, analysts have to choose which 'distance metric to use in order to cluster objects, and although there is no definitive way of choosing the 'right' distance metric, the choice can have profound effects on how individuals or conditions cluster.

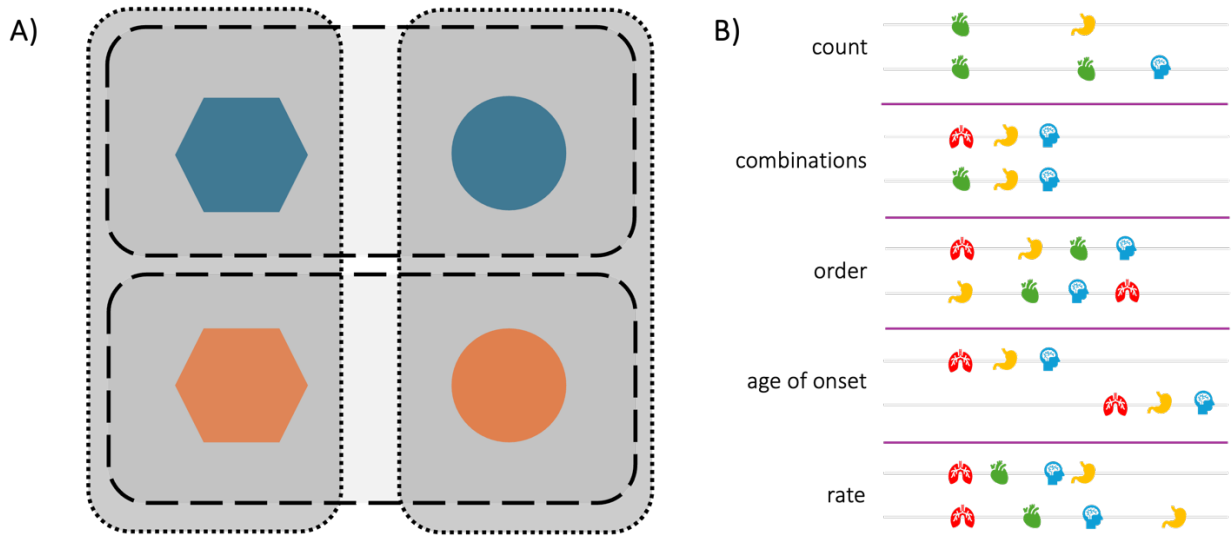

*Figure 2: A. **Non-unique clustering:** Examples of non-uniqueness of clustering. Here 4 objects can be divided in two clusters based on either their shapes (dotted boxes) or their colours (dashed boxes). Both solutions are valid. B. **Different characteristics of multimorbidity trajectories:** The figure illustrates the various ways multimorbidity trajectories can differ. Here, the trajectories in the latter rows differ from the first rows in terms of the count, combinations, order, onset and rate of conditions being accrued, and different similarity measures can account for these different facets of the trajectories.*

Why cluster multimorbidity?

There are a number of situations where a clustering approach is potentially useful, but these purposes are not always explicitly stated in published papers, nor is it always clear whether methodological choices are aligned with the intended purpose.

*Understanding aetiology and mechanisms*

Understanding how conditions cluster can inform consideration of aetiology and the mechanisms underlying the co-occurrences. Uncovering mechanisms has the potential to target further discovery science to better delineate downstream pathways, and ultimately design interventions at an individual or population level. For example, understanding the genetics underpinning different patterns of morbidity accrual can facilitate drug discovery by targeting appropriate pathways[3].

*Natural history and predicting adverse outcomes*

Understanding the historical and predicted pattern of multimorbidity accrual over an individual's lifetime may be useful for targeting interventions and encouraging changes in life choices early. Similarly, historical patterns of long-term conditions may be predictive of a wide range of future adverse outcomes (mortality, functional impairment, reduced quality of life, loss of independence, risk of particular conditions like cardiovascular disease), and such prediction tools have an obvious potential use for population risk stratification to target interventions[7].

*Planning health and social care service*

Understanding which long-term conditions co-occur frequently may help with better service planning to reduce the individual and economic impact of these clusters. For example, diabetes is common in pregnancy and associated with adverse outcomes, and cardiovascular disease commonly occurs in people with severe mental illness and significantly contributes to high premature mortality. Irrespective of whether there is any aetiological connection, understanding co-occurrence 'clusters' can usefully inform service planning and delivery.

Methodological choices

Different intended purposes of the multimorbidity clusters might lead to different choices of clustering. For example, understanding mechanisms requires the conditions in the cluster to be associated through an underlying common cause, predicting adverse outcomes requires the conditions in the cluster to influence each other and the outcome, and planning health and social care services requires the conditions to be highly prevalent or highly impactful on individuals or services. While there is little evidence that any one choice is better than any other[8], there are some critical choices to highlight.

Choosing which conditions to include, and how to ascertain them

In the multimorbidity field in general, there is a very large variation between studies in terms of which conditions they choose to analyse and how they ascertain those conditions. A systematic review of 566 studies, the median number of conditions included was 17 (interquartile range 11-23) and only eight conditions (and no mental health conditions) were included in more than half of the studies[9]. Clustering studies show similar variation, with almost half of 172 studies of multimorbidity patterns only including 10-20 conditions (the review excluded studies with <10 conditions), and only 10% including >100[8]. The clusters observed in data will be influenced by the choice of conditions to analyse, which typically favours cardiovascular conditions, diabetes and cancer over mental health, neurological and musculoskeletal conditions[9].

More recent studies examining clustering often include a large number of conditions, for example, defined by three-digit ICD-10 codes[8,10] but it is important to recognise that these may not always uniquely identify distinct 'conditions' in the clinical sense (for example, there are multiple three-digit ICD-10 codes defining various manifestations of type 1 diabetes), meaning that spurious associations may be present. There is no correct answer to the question as to the optimal level of granularity, but decisions made are non-trivial (Box 1). More granular definitions risk spurious 'overlap' but coarser definitions can create overly heterogenous 'conditions'. As well as choice of conditions, researchers have to choose how to ascertain them, including which datasets to use, and how to define each condition within the chosen dataset(s).

*Datasets* vary considerably in terms of their size, what data they have (both condition-related and other attributes), and the period over which data are available (coverage). Space precludes a detailed discussion, but in multimorbidity research, the choice is often between longitudinal routine datasets based on administrative (or electronic health record data), and bespoke research cohorts. Routine data can be very large, but often will not provide a true lifetime history because they are bounded by the start of electronic data recording or the introduction of unique identifier that allows the linkage of episodes to individuals (and so are often not available in low and some middle-income countries). Routine data are often good for recording conditions and basic demography, but have considerable missing data for other variables like smoking, alcohol, ethnicity, social circumstances and symptoms. In contrast, research cohorts may have longer (and sometimes near lifetime) coverage and typically have more detailed information about individual circumstances and social context that is absent from routine data, but conversely may have less granular data about health conditions and may be too small for many clustering purposes (or at least unable to account for rare conditions). More recently, routine data linkage to research cohorts provides the benefits of both, particularly for new very large research cohorts such as UK Biobank, although researchers need to think carefully about the extent and quality of the linkage and the coverage of the data being linked[11].

*Condition ascertainment* will be affected by choice of dataset, choice of codelist and other rules to apply, and whether what is being measured is lifetime history or 'active disease'. Choices of the data source will influence the ascertainment of conditions, with inpatient data alone having low sensitivity compared to primary care data for conditions which are rarely the direct cause of admission, notably mental health and common musculoskeletal conditions[12]. Increasingly, reseachers use pre-existing lists of diagnostic codes to define the presence of conditions[13], but validation is variable and codelists are rarely identical. For some conditions like chronic kidney disease, ascertainment may be better done using laboratory data or combinations of laboratory and coded data, or it may be appropriate to apply a set of rules on top of any codelist used, for example to define type 1 and type 2 diabetes. Additional rules may also be useful for conditions like asthma, epilepsy and depression which often remit but sometimes relapse. For genetic studies, the lifetime history of each condition may be appropriate, whereas for prediction, 'active condition' may be more appropriate. Finally, when doing longitudinal analysis, changes in how disease is conceptualised or defined may affect results, with chronic kidney disease being an example where biochemical definitions based on estimated glomerular filtration rate had a large impact in the early 2000s.

Irrespective of how individual conditions are ascertained, all data is subject to varying observability. Historical data is likely to be less complete and less granular than more recent data, and dates of diagnosis in routine data may also be approximate or inaccurate, which will constrain approaches which examine sequences or rates of condition accrual more than those which examine combinations or counts. All data is only a record of what has happened so far, which is not problematic for prediction, but if clustering is intended to inform aetiological understanding, then the full phenotype may not yet have been observed in younger age groups. All data are also subject to right censoring of various kinds, including loss to follow-up in research cohorts, deregistration from the clinical services providing routine data, and competing mortality that is often not explicitly addressed or acknowledged in the literature.

There is, therefore, a need for greater consensus on which conditions to include (and how to ascertain them) and greater transparency from researchers as to why they have chosen a particular set of conditions measured with a particular level of granularity over a particular period (during which observability may vary) with a particular type of definition (such as lifetime history versus recently active). There is no single correct set of choices to be made, but there is consensus on a core group of conditions to include in multimorbidity research, and we encourage researchers to consider this when designing studies (and to vary from it when that is appropriate, with explicit justification)[14].

Box 1: Example levels of granularity of conditions in ICD-10

| Element of ICD-10 hierarchy[a] | Level of granularity | Lifelong versus active disease |
|---|---|---|
| J40-47 Chronic lower respiratory diseases<br>  J40 Bronchitis, not specified as acute or chronic<br>  J41 Simple and mucopurulent chronic bronchitis<br>  J42 Unspecified chronic bronchitis<br>  J43 Emphysema<br>  J44 Other chronic obstructive pulmonary disease<br>  J45 Asthma<br>  J46 Status asthmaticus<br>  J47 Bronchiectasis | Higher level 'standard' group is heterogenous<br><br>J40-J44 are variably included in definitions of chronic obstructive pulmonary disease (COPD)[b]<br><br>J45 and J46 are both asthma (status asthmaticus is a very severe asthma attack)<br>J47 is distinct from both asthma and COPD | COPD is lifelong from diagnosis<br><br>Asthma commonly remits[c] |
| J45 Asthma<br>  J45.0 Predominately allergic asthma<br>  J45.1 Nonallergica asthma<br>  J45.8 Mixed asthma<br>  J45.0 Asthma, unspecified | Lower levels of the ICD-10 hierarchy may define condition subtypes which may be distinct (e.g., genetically, in relation to prognosis), but usually contain an 'unspecified' category which may be commonly used | |

a. Other coding systems are much more highly elaborated (eg Read v2 in UK primary care; SNOMED-CT across UK healthcare in the future)

b. For example, see https://phenotypes.healthdatagateway.org/phenotypes/PH43/version/86/detail/ and https://phenotypes.healthdatagateway.org/phenotypes/PH798/version/2222/detail/

c. Researchers more interested in applied health services research often measure 'active asthma' by using recent prescriptions to define asthma as active[15]. Since the same medications are used in COPD (and because of the historical miscoding of COPD as asthma, this may mean disallowing the co-existence of COPD and asthma in that context (although overlap syndromes do exist). Different choices may be better in aetiological studies where for example lifetime history of asthma may be more useful than active asthma in genetic studies.

Cross-sectional versus longitudinal clustering

Clustering can either use cross-sectional or longitudinal data. In cross-sectional analysis, the focus is usually on the count or the combination of conditions present at a particular moment in time. Longitudinal analysis can also account for other characteristics, such as the order in which the conditions are accrued, the age of onset at which conditions are accrued, or the rate of accrual. Cross-sectional methods have been better explored than longitudinal methods, with a recent systematic review observing that existing methods for longitudinal clustering "characterize accumulation and disease combinations, and to a lesser extent disease sequencing and progression"[16], and although, a wider range of methods have been proposed for longitudinal clustering[3] more recently, what is possible may be constrained by the data available. Both cross-sectional and longitudinal approaches can be useful in various applications, and we suggest that the choice of one domain over the other should be adequately justified in the context of downstream analysis, and the availability of data and computational resources.

Clustering conditions versus clustering individuals

Clustering long-term conditions can either involve grouping individuals based on the conditions they have ('clustering individuals') or grouping conditions based on how they co-occur in individuals ('clustering conditions'). Similarities between individuals can be computed in various ways (for example, by observing the number of long-term conditions shared between them), while similarities between conditions can be computed using a measure of association (such as correlation coefficient or relative risk). Although it is possible to apply findings from one level at the other level (clustering of individuals can inform understanding of condition clustering and vice versa), clustering conditions may lend itself better to aetiological research, whereas clustering individuals may be more useful for service planning or prediction.

Exclusive versus non-exclusive assignment

Clustering commonly involves grouping objects like conditions or individuals into exclusive sets, but this can be restrictive[17]. When clustering conditions, we often have good reasons to expect that conditions could belong to multiple groups. Type 2 diabetes is much more common in people with obesity, so it would be expected to cluster with other obesity-related conditions like arthritis. Type 2 diabetes is also a cause of atherosclerotic cardiovascular diseases, so it would be expected to cluster with these. If there are multiple mechanisms by which conditions cluster, then non-exclusive allocation to clusters is plausibly more appropriate. For example, our existing understanding is that although we might observe that an individual has all of 'diabetes, asthma, coronary heart disease (CHD), eczema, hypertension, hay fever', there are actually groups of related conditions - cardiometabolic (diabetes, CHD, hypertension) and atopy (asthma, eczema, hay fever), rather than all six conditions being related, and here it is also possible to consider non-exclusive allocation when clustering individuals, i.e., an individual may belong to both cardiometabolic and atopy clusteres. We may argue that, from the perspective of facilitating treatment decisions or allocation to care pathways, exclusive assignment may be more desired, while from the perspective of understanding aetiology, non-exclusive assignment may be more helpful.  illustrates these different clustering approaches.

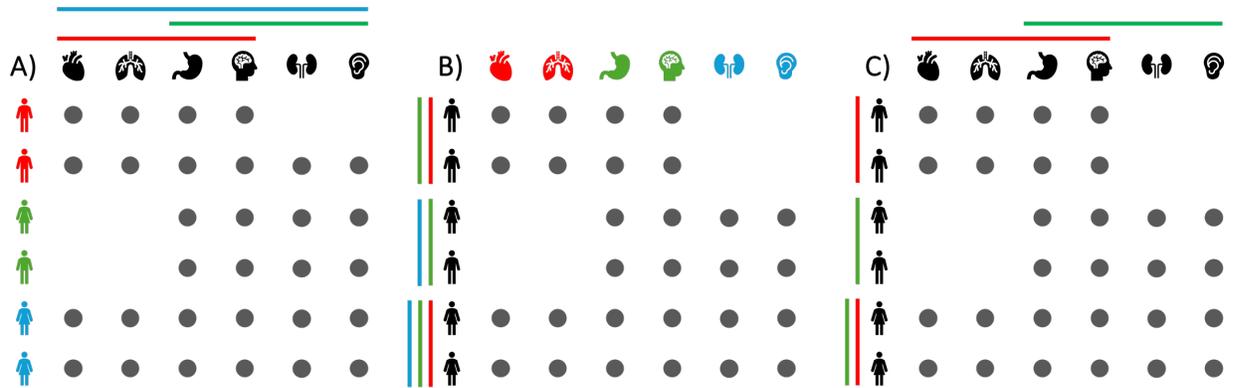

*Figure 3: A) Exclusive clustering of individuals, while the same condition can be assigned to multiple clusters. B) Exclusive clustering of conditions, while the same individual can be assigned to multiple clusters, C) Non-exclusive assignment of both conditions and individuals.*

Prevalence versus association

Existing clustering approaches can be broadly divided into two groups: prevalence-based and association-based. Existing clustering approaches mostly focus on prevalence, i.e., combinations, sequences or other patterns that appear more frequently than others are identified as clusters, even though high frequency may partly or wholly be chance co-occurrence of common conditions rather than relatedness among the conditions. Association-based approaches, on the other hand, focus on capturing such relatedness explicitly. We argue that both these approaches can be important. From the perspective of care delivery, frequent co-occurrence (even if due to chance) may be highly relevant for planning services, and for prediction or prognosis if co-occurrence is strongly associated with outcomes. From the perspective of understanding aetiology or identifying shared mechanisms, association beyond chance may be a more relevant perspective.

What are the desired characteristics of clustering models?

Multimorbidity clustering solutions will vary depending on the kinds of choices of dataset, conditions to include, condition ascertainment, and method of clustering outlined above. There is no current way to definitively conclude that one cluster solution is more valid than another. Dhafari et al identified five ways in which 111 existing studies sought to validate cluster solutions (some used more than one method), namely: examining association with clinical outcomes (n=79), stability across subsamples (n=26), clinical plausibility (n=22), stability across methods (n=7), and exploring common determinants (n=2)[8]. The types of 'validity' being examined with these methods are rather different, but this does provide a basis for a more elaborated list of desirable attributes of analyses clustering multimorbidity (Box 2).

Box 2: Desired characteristics of analyses clustering multiple long-term conditions

| Desired characteristics of clustering long-term conditions | |
|---|---|
| Stability to perturbation | Inferred clusters should be robust to small perturbation in the data. Assessing stability is informative particularly for conditions of low prevalence since any inferred clusters may be sensitive to noise in the data (from miscoding or varying observability). |
| Stability across methods | Inferred clusters should ideally be similar when different methods are used in the same dataset. Although the assumptions made by different methods might lead to different clustering, we would expect that clustering approaches that focus on the same attribute (for example, combination or order) to find similar clusters. |
| Stability across datasets (reproducibility) | Reproducibility implies that the inferred clusters in different datasets developed using the same methods are similar. This provides confidence in the inferred clusters at a population level since the datasets might not be representative of the whole population, and thus, not all clusters found can be similar but we expect the prominent clusters to replicate. |
| Explainability | Explainability[18] implies that there is an adequate human-level understanding of how the cluster is being formed at the group level (and the characteristics of clusters), and how an individual is being assigned to the group at an individual level. For studies involving large numbers of conditions, this is far from straightforward to achieve. |
| Fairness | Fairness implies that, at an individual level, an individual should be allocated to an appropriate cluster nearby cluster (**Figure 4**), and at a group level, the clusters should not be biased inadvertently to a specific demographic that might be a sensitive characteristics. |
| Clinical plausibility or utility and mechanistic underpinnings (operationalizability) | Operationalizability implies that the inferred clusters are appropriate for relevant downstream analysis, e.g., predicting risk or understanding drivers (**Figure 5**). Clinical plausibility based on expert judgement is commonly used to 'validate' cluster solutions, but is problematic since it is often based on one or two chosen exemplars rather than the full model output, and new and unexpected clusters are by definition not (yet) clinically plausible. If inferred clusters are associated with outcomes like mortality, emergency hospital admission and institutionalisation, then this is often taken to be evidence the cluster is valid, but it is unlikely that any condition count will not be so associated (because 'sicker people are sicker'). Similarly, evidence of shared genetic or metabolic underpinnings for conditions which cluster, or evidence that similar treatments have benefits across conditions in inferred clusters would support the validity of inferred clusters. |

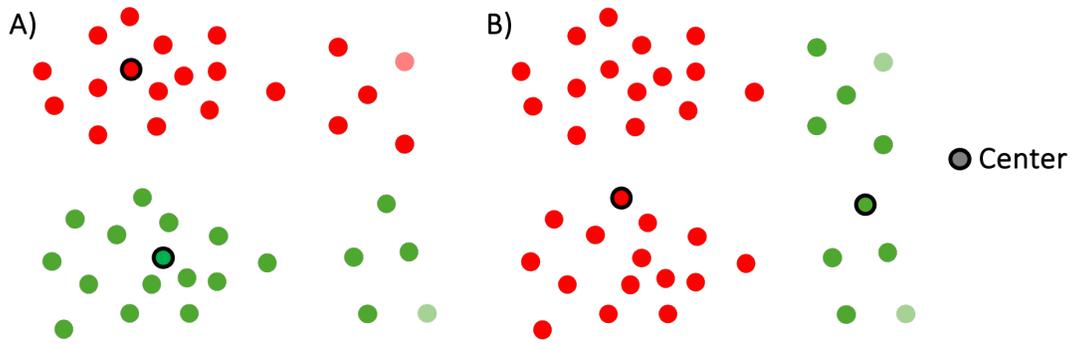

*Figure 4: A) Unfair clustering of individuals since an individual (shown as light color) can be far away from the respective cluster centre, B) Fair clustering since every individual is only a certain distance away from the respective cluster centre.*

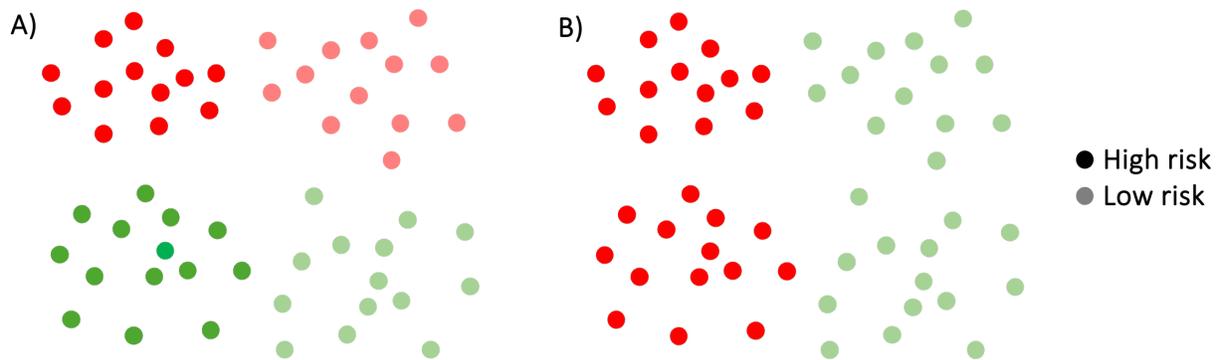

*Figure 5: A) Less operationalizable clustering since both high and low risk individuals can belong to the same cluster, B) more operationalizable clustering since it separates the high and low risk individuals.*

## Conclusion

Multimorbidity is a growing challenge to health and care systems globally. Research focussing on multimorbidity clusters has the *potential* to uncover underpinning mechanisms and better target interventions to improve outcomes, although we believe there is little evidence that this approach has been fruitful as yet. Multimorbidity clustering approaches must account for the scarcity and quality of data, be mindful of the (usually implicit) choices being made, and the downstream analysis being catered for. There is no clearly superior way to approach this kind of analysis, but a clearer description and justification of the choices made will improve comparability and reproducibility of what is currently a diverse and difficult to synthesise field. We encourage moving from a *top-down* approach of unsupervised clustering of long-term conditions and exploring generic attributes of the clusters to a *bottom-up* approach of considering the purpose of clustering in a particular context, using this to guide the choices made during clustering analysis, and reporting and explicitly justifying these choices to improve comparability and reproducibility, and facilitate evaluation of utility for different purposes.

## Acknowledgment

This study/project is partly funded by the National Institute for Health Research (NIHR) Artificial Intelligence and Multimorbidity: Clustering in Individuals, Space and Clinical Context (AIM-CISC) grant NIHR202639. The views expressed are those of the author(s) and not necessarily those of the NIHR or the Department of Health and Social Care. In addition, Seth's and Guthrie's role in this research is partly funded by the Legal & General Group (research grant to establish the independent Advanced Care Research Centre at the University of Edinburgh). The funder had no role in the conduct of the study, interpretation or the decision to submit for publication. The views expressed are those of the authors and not necessarily those of Legal & General.

# References

1. Whitty, C. J. M. & Watt, F. M. Map clusters of diseases to tackle multimorbidity. *Nature* **579**, 494–496 (2020).

2. Whitty, C. J. M. *et al.* Rising to the challenge of multimorbidity. *BMJ* l6964 (2020) doi:10.1136/bmj.l6964.

3. Jiang, X. *et al.* Age-dependent topic modeling of comorbidities in UK Biobank identifies disease subtypes with differential genetic risk. *Nat. Genet.* (2023) doi:10.1038/s41588-023-01522-8.

4. Makovski, T. T., Schmitz, S., Zeegers, M. P., Stranges, S. & Van Den Akker, M. Multimorbidity and quality of life: Systematic literature review and meta-analysis. *Ageing Res. Rev.* **53**, 100903 (2019).

5. Soley-Bori, M. *et al.* Impact of multimorbidity on healthcare costs and utilisation: a systematic review of the UK literature. *Br. J. Gen. Pract.* **71**, e39–e46 (2021).

6. The Academy of Medical Sciences, Medical Research Council, ., National Institute of Health Research, . & Wellcome Trust, . Advancing research to tackle multimorbidity: the UK and LMIC perspectives. in (London, 2018).

7. Arakelyan, S. *et al.* Effectiveness of holistic assessment–based interventions in improving outcomes in adults with multiple long-term conditions and/or frailty: an umbrella review protocol. *JBI Evid. Synth.* **21**, 1863–1878 (2023).

8. Dhafari, T. B. *et al.* A scoping review finds a growing trend in studies validating multimorbidity patterns and identifies five broad types of validation methods. *J. Clin. Epidemiol.* **165**, 111214 (2024).

9. Ho, I. S.-S. *et al.* Examining variation in the measurement of multimorbidity in research: a systematic review of 566 studies. *Lancet Public Health* **6**, e587–e597 (2021).

10. Jensen, A. B. *et al.* Temporal disease trajectories condensed from population-wide registry data covering 6.2 million patients. *Nat. Commun.* **5**, 4022 (2014).

11. Prigge, R. *et al.* Robustly Measuring Multiple Long-Term Health Conditions Using Disparate Linked Datasets in UK Biobank. Preprint at https://doi.org/10.2139/ssrn.4863974 (2024).

12. MacRae, C. *et al.* Impact of data source choice on multimorbidity measurement: a comparison study of 2.3 million individuals in the Welsh National Health Service. *BMC Med.* **21**, 309 (2023).

13. Kuan, V. *et al.* A chronological map of 308 physical and mental health conditions from 4 million individuals in the English National Health Service. *Lancet Digit. Health* **1**, e63–e77 (2019).

14. Ho, I. S. S. *et al.* Measuring multimorbidity in research: Delphi consensus study. *BMJ Med.* **1**, e000247 (2022).

15. Barnett, K. *et al.* Epidemiology of multimorbidity and implications for health care, research, and medical education: a cross-sectional study. *The Lancet* **380**, 37–43 (2012).

16. Cezard, G., McHale, C., Sullivan, F., Bowles, J. & Keenan, K. *Studying Trajectories of Multimorbidity: A Systematic Scoping Review of Longitudinal Approaches and Evidence*. http://medrxiv.org/lookup/doi/10.1101/2020.11.16.20232363 (2020) doi:10.1101/2020.11.16.20232363.

17. Frank, M., Streich, A. P., Basin, D. & Buhmann, J. M. Multi-Assignment Clustering for Boolean Data. *J. Mach. Learn. Res.* **13**, 459–489 (2012).

18. Molnar, C. *Interpretable Machine Learning: A Guide for Making Black Box Models Explainable*. (2025).